\documentclass[aps,prb,twocolumn,showpacs,superscriptaddress,groupedaddress]{revtex4-1}

\usepackage{amsmath}
\usepackage{amssymb}   
\usepackage{graphicx}
\usepackage{CJK}

\begin{document}

\title{Effect of interaction on the Majorana zero modes in the Kitaev chain at half filling}

\author{Zhidan Li$^1$}
\author{Qiang Han$^{1,2}$}
\address{$^1$Department of Physics, Renmin University of China, Beijing 100872, China}
\address{$^2$Beijing Key Laboratory of Opto-electronic Functional Materials and Micro-nano Devices, Renmin University of China, Beijing 100872}

\date{\today}

\begin{abstract}

The one dimension interacting Kitaev chain at half filling is studied. The symmetry of the Hamiltonian is examined by dual transformations and various physical quantities as functions of the fermion-fermion interaction $U$ are calculated systematically using the density matrix renormalization group method. A special value of interaction $U_p$ is revealed in the topological region of the phase diagram. We show that at $U_p$ the ground states are strictly two-fold degenerate even though the chain length is finite and the zero-energy peak due to the Majorana zero modes is maximally enhanced and exactly localized at the end sites. $U_p$ may be attractive or repulsive depending on other system parameters. We also give a qualitative understanding of the effect of interaction under the self-consistent mean field framework. \\


\end{abstract}


\maketitle

In recent years, the topological states of matter, such as topological insulators, topological superconductors and topological semimetals etc, have attracted enormous interest in the condensed matter community on the topic of the topologically protected boundary states. Especially, a one dimensional (1D) model of $p$-wave topological superconductor was proposed by Kitaev,~{\cite{Ref_Kitaev_0} which hosts an unpaired Majorana zero mode (MZM) at each of its two ends. Although the $p$-wave superconductors are rare in nature, various 1D artificial topological superconducting systems have been proposed theoretically~\cite{Ref_Fu_Kane,Ref_Semi_0,Ref_Semi_1} to realize effective $p$-wave pairing in real materials.
Later several experimental groups reported signatures of observing the MZMs~\cite{Ref_Experiments_0,Ref_Experiments_1,Ref_Experiments_2,Ref_Experiments_3,Ref_Experiments_4,Ref_Experiments_5} in the spin-orbit coupled semiconductor nanowires in proximity to an $s$-wave superconductor.

However, in Kitaev's seminal paper~{\cite{Ref_Kitaev_0} the existence of MZMs was revealed in the single particle picture and the many body effect due to the fermion-fermion interaction was not addressed thoroughly. Theoretical works on the interacting Kitaev-like chain~\cite{Ref_interaction_0,Ref_interaction_1,
Ref_interaction_2,Ref_interaction_3,Ref_interaction_4,Ref_interaction_5,Ref_interaction_6,Ref_interaction_7,Ref_interaction_8} found that the topological superconducting phase is
susceptible to the fermion interaction which can either close the superconducting gap or induce competing orders. It was found that the interacting Kitaev chain will enter into a trivial (incommensurate) charge density wave phase (CDW) or Schr\"{o}dinger-cat-like phase~\cite{Ref_interaction_8} in the region of strong repulsive or attractive interactions.~\cite{Ref_interaction_2,Ref_interaction_3,Ref_interaction_5,Ref_interaction_8} As for the effect of interaction on the MZMs, the numerical analysis~\cite{Ref_interaction_3} based on the density matrix renormalization group (DMRG) method showed that the MZMs will survive under moderate interactions and the Majorana zero-bias peak is enhanced (weakened) by weak attractive (repulsive) interaction.

Here in this paper we will clarify the effect of the nearest neighboring interaction $U$ on the MZMs in the Kitaev chain using the dual transformations, the DMRG method and the self-consistent mean field (SCMF) method. We focus on the half filling case to avoid the complication due to the coexistence of superconducting order with the incommensurate CDW phase. We find a special value of interaction $U_p$ where the Majorana zero-energy peak is exactly localized even for a finite-size chain.
If $U$ is less or more than $U_p$, the MZMs will be suppressed, no matter whether the interaction is attractive or repulsive.

The Hamiltonian of the 1D interacting Kitaev chain is written as,
\begin{align}\label{Eq_Ham_00}
    H = & \sum_{j=1}^{L-1} \big[ (- t c_{j}^{\dagger} c_{j+1} - \Delta c_{j}^{\dagger}c_{j+1}^{\dagger} + \text{H.c.}) \nonumber \\
       & \left. + U \left( n_j-\frac{1}{2} \right) \!\! \left( n_{j+1}-\frac{1}{2} \right) \right]
        -\mu \sum_{j=1}^{L} n_j,
\end{align}
where $c_{j}^{\dagger}$ denotes the fermion creation operator on site $j$, $n_j=c_j^\dagger c_j$ is the fermion number operator, and $L$ is the length of the chain. $t$ is the nearest-neighbor hopping integral, $\Delta$ the $p$-wave pairing potential, $\mu$ the chemical potential, and $U$ the nearest-neighbor interaction.
Without loss of generality, $t$ and $\Delta$ are set as real and positive. In this paper, we only consider the case $\mu=0$, and accordingly the system is always at half filling; and $t$ is chosen as unit of energy.

For the noninteracting case ($U=0$), the system at half filling is in the topological superconducting phase and there exists one unpaired MZM localized at each end of the chain. The corresponding zero-energy peak decays exponentially and the characteristic length dependens on $t$ and $\Delta$. For $t=\Delta$ the MZM is exactly localized as shown by Kitaev.~ \cite{Ref_Kitaev_0} In the presence of the $U$ term, there is a topological region in the parameter space of $U$, whose boundaries can be given rigorously after transforming Eq.~(\ref{Eq_Ham_00})
into an integrable spin $XYZ$ model~\cite{Ref_Baxter} as shown below.

Through a Jordan-Wigner transformation,
\begin{equation}\label{Eq_J-W}
c_j^{\dagger} = \sigma_j^{+} \displaystyle \prod_{k=1}^{j-1} (-\sigma_k^{z}),\ \ \ c_j = \sigma_j^{-} \displaystyle \prod_{k=1}^{j-1} (-\sigma_k^{z}),
\end{equation}
the interacting Kitaev chain can be mapped to the Heisenberg 1D $XYZ$-model,
\begin{align}\label{Eq_Ham_XYZ}
    H \!=\! -\frac{1}{2} \!\sum_{j=1}^{L-1} \!\left[ \!(t+\Delta) \sigma_{j}^{x}\sigma_{j\!+\!1}^{x} + \!(t-\Delta)  \sigma_{j}^{y}\sigma_{j\!+\!1}^{y}
                                            - \!\frac{U}{2} \sigma_{j}^{z}\sigma_{j\!+\!1}^{z} \right]\!.
\end{align}

\begin{figure}[h]
\begin{center}
  \includegraphics[width=8.0cm, angle=0]{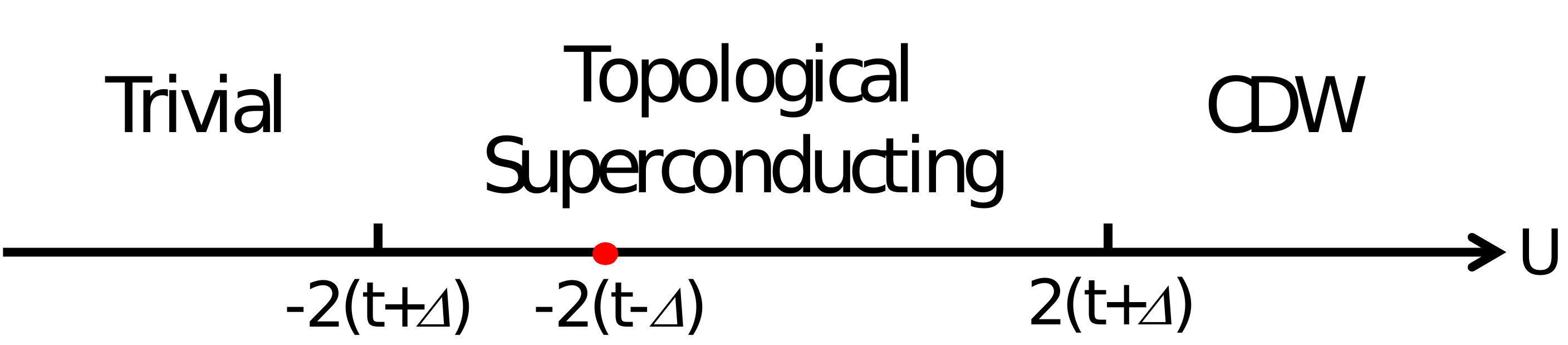}
\end{center}
 \caption{(Color online) Phase diagram of 1D interacting Kitaev model at half filling.
         }
 \label{Fig_PD_XYZ}
\end{figure}
Eq.~(\ref{Eq_Ham_XYZ}) has three highly symmetric self-dual points on the axis of $U$, namely $\pm U_c=\pm2(t+\Delta)$ and $U_p=-2(t-\Delta)$. To be specific, when $U=U_c$, Eq.~\eqref{Eq_Ham_XYZ} is invariant under the dual transformation
$\sigma_{j}^{x} \leftrightarrow (-1)^{j}\sigma_{j}^{z}, \sigma_{j}^{y} \leftrightarrow \sigma_{j}^{y}$. Likewise when $U=-U_c$, the $XYZ$ model is invariant under
$\sigma_{j}^{x} \leftrightarrow \sigma_{j}^{z}, \sigma_{j}^{y} \leftrightarrow \sigma_{j}^{y}$. $\pm U_c$ are two critical points seperating three different phases: the trivial phase $U<-U_c$, the CDW phase for $U>U_c$ and the topological superconducting phase (TSC) in between ($-U_c<U<U_c$) as showed in Fig.~\ref{Fig_PD_XYZ}, which is consistent with previous works.~\cite{Ref_interaction_2,Ref_interaction_3,Ref_DP_of_XYZ_0,Ref_DP_of_XYZ_1,Ref_DP_of_XYZ_2} The third self-dual point $U_p=-2(t-\Delta)$ is also depicted in Fig.~\ref{Fig_PD_XYZ}, with the corresponding dual transformation $\sigma_{j}^{x} \leftrightarrow \sigma_{j}^{x}, \sigma_{j}^{y} \leftrightarrow \sigma_{j}^{z}$. However different from $\pm U_c$, $U_p$ does not indicate a phase transition but a turning point at which we find that the ground states exhibit exactly double degeneracy even for a finite-size chain and the MZM is maximally localized at just the outmost lattice site as discussed below.

In the following we further address the phase diagram and the effect of $U$ on the topological properties via the DMRG method.~\cite{Ref_White_0,Ref_White_1,Ref_Schollwock} The following quantities are obtained: (1) the many-body energy spectrum; (2) the entanglement entropy; (3) the local density of states (LDOS) as functions of energy and position. The entanglement entropy is defined as $S=-\text{Tr}({\rho^r \ln\rho^r})$,
where $\rho^r$ is the reduced density matrix. The LDOS can be obtained from the retarded Green's function according to,
\begin{equation}\label{}
    \rho(j,\omega) = -\frac{1}{\pi} \text{Im} G^{\text{R}}(j,\omega),
\end{equation}
where
\begin{align} \label{green}
    G^{\text{R}}(j,\omega) & =    \langle \psi_0 | C_j~ \frac{1}{\omega+i\eta +E_{0}-\hat{H}} ~C_j^{\dagger} | \psi_0 \rangle \nonumber \\
                    & +   \langle \psi_0 | C_j^{\dagger}~ \frac{1}{\omega+i\eta-E_{0}+\hat{H}} ~C_j | \psi_0 \rangle,
\end{align}
with $\eta\rightarrow0^+$. $E_0$ and $|\psi_0\rangle$ denotes the ground-state energy and vectors of $\hat H$.
The LDOS can be calculated by a hybrid method of DMRG and kernel polynomial method.~\cite{Ref_interaction_3,Ref_Schollwock,Ref_KPM} In the numerical investigation the length of chains $L$ is around 20 $\sim$ 40 sites, and the number of states kept in DMRG is 500$\sim$2000.

\begin{figure}[ht]
\begin{center}
  \includegraphics[width=8.0cm, angle=0]{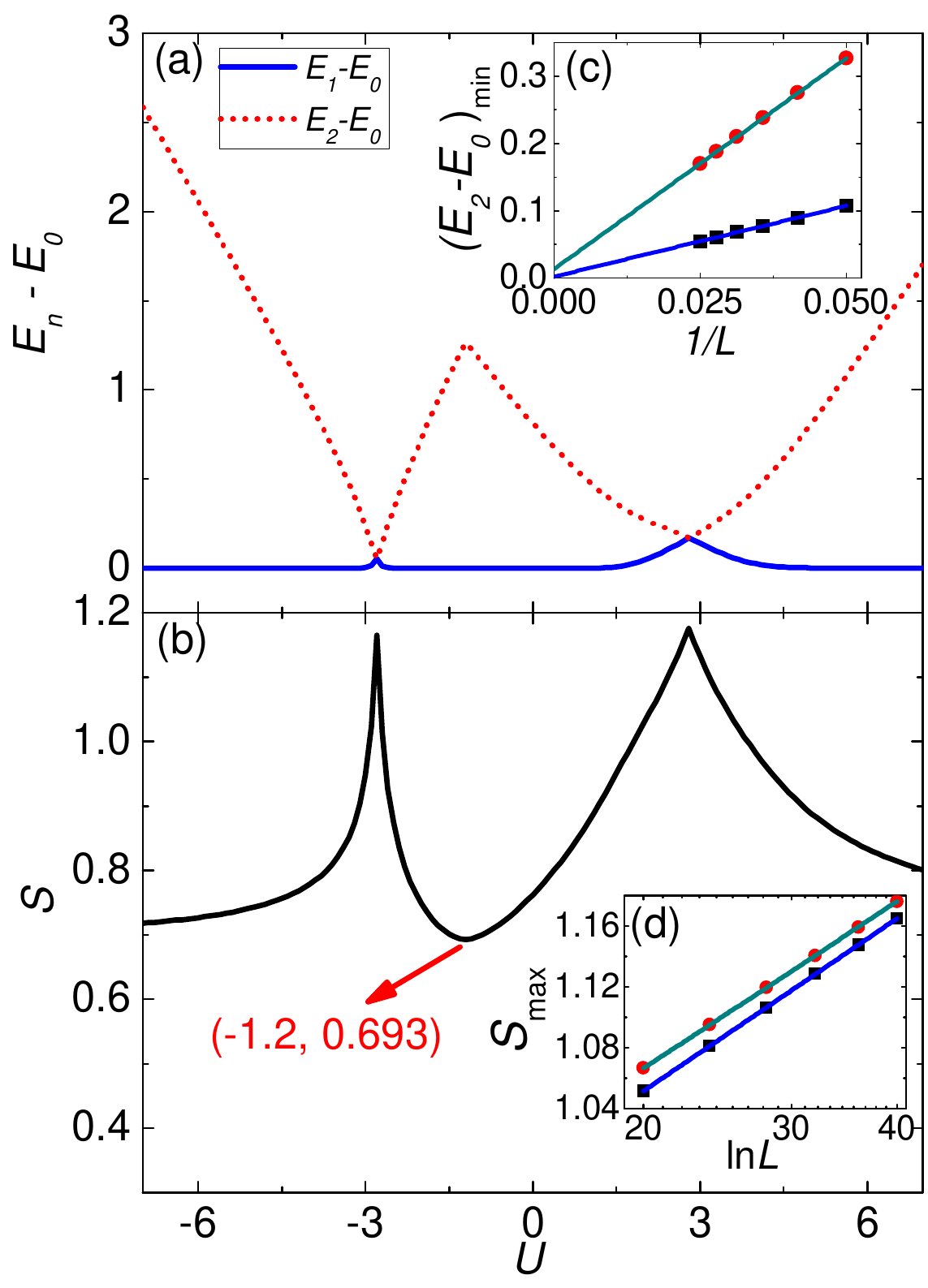}
\end{center}
 \caption{(Color online)
 (a) Many body low-energy excitation spectrum of the 1D interacting Kitaev chain as a function of $U$.
 Blue solid and red dotted lines denote the first ($E_1$) and second ($E_2$) excitation energies relative to the ground-state energy ($E_0$), respectively.
 (b) The entanglement entropy ($S$) of the 1D interacting Kitaev chain as a function of $U$.
 (c) Inset of (a) shows the finite-size scaling of the two local minima of $E_2-E_0$.
 (d) Inset of (b) shows the finite-size scaling of the two local maxima of $S$.
 The system parameters are $\Delta$=0.4 and $L=40$.}
 \label{Fig_En}
\end{figure}
Fig.~\ref{Fig_En} shows the $U$-dependent many body energy spectrum and entanglement entropy of a finite-size chain under the open boundary condition.
It can be found that the ground states are two-fold degenerate in all three phases by the finite-size scaling analysis with $E_1-E_0\to 0$ when $L\rightarrow\infty$, which is consistent with the exact integrable $XYZ$ model~\cite{}.
The finite-size scaling (Fig.~\ref{Fig_En}(c)) shows that the excitation gap $E_2 - E_0$ is closed at $U=\pm 2.8$ when $L\rightarrow\infty$, which signifies two phase transition points identical to the self-dual points $\pm U_c=\pm2(t+\Delta)=\pm 2.8$ for $\Delta=0.4$.
The two phase transitions can be further captured by the entanglement entropy which shows divergent behavior at $\pm U_c$ as shown in Fig.~\ref{Fig_En}(b).
By the way, the finite-size scaling (Fig.~\ref{Fig_En}(d)) finds that the entanglement entropy $S$ at $\pm U_c$ has a relation with the chain length $L$ as
$S=\frac{c}{6}\text{ln}L+\text{const}$ and $c\thickapprox1$, which is in agreement with the results of the conformed field theory, and the universal central charge $c\thickapprox1$ is a typical value of the 1D Heisenberg model.~\cite{Ref_CFT}
In the topological region $-U_c<U<U_c$, Fig.~\ref{Fig_En} also shows a special point $U_p=-2(t-\Delta)=-1.2$ for $\Delta=0.4$, where the excitation gap takes a maximum value while the entanglement entropy has a minimum value ($S=\text{ln}2\approx0.693$ at $U=-1.2$ for any chain length). This results motivate us to further investigate the effect of interaction on MZM by studying the variation of LDOS with $U$.

\begin{figure}[ht]
\begin{center}
  \includegraphics[width=7.5cm, angle=0]{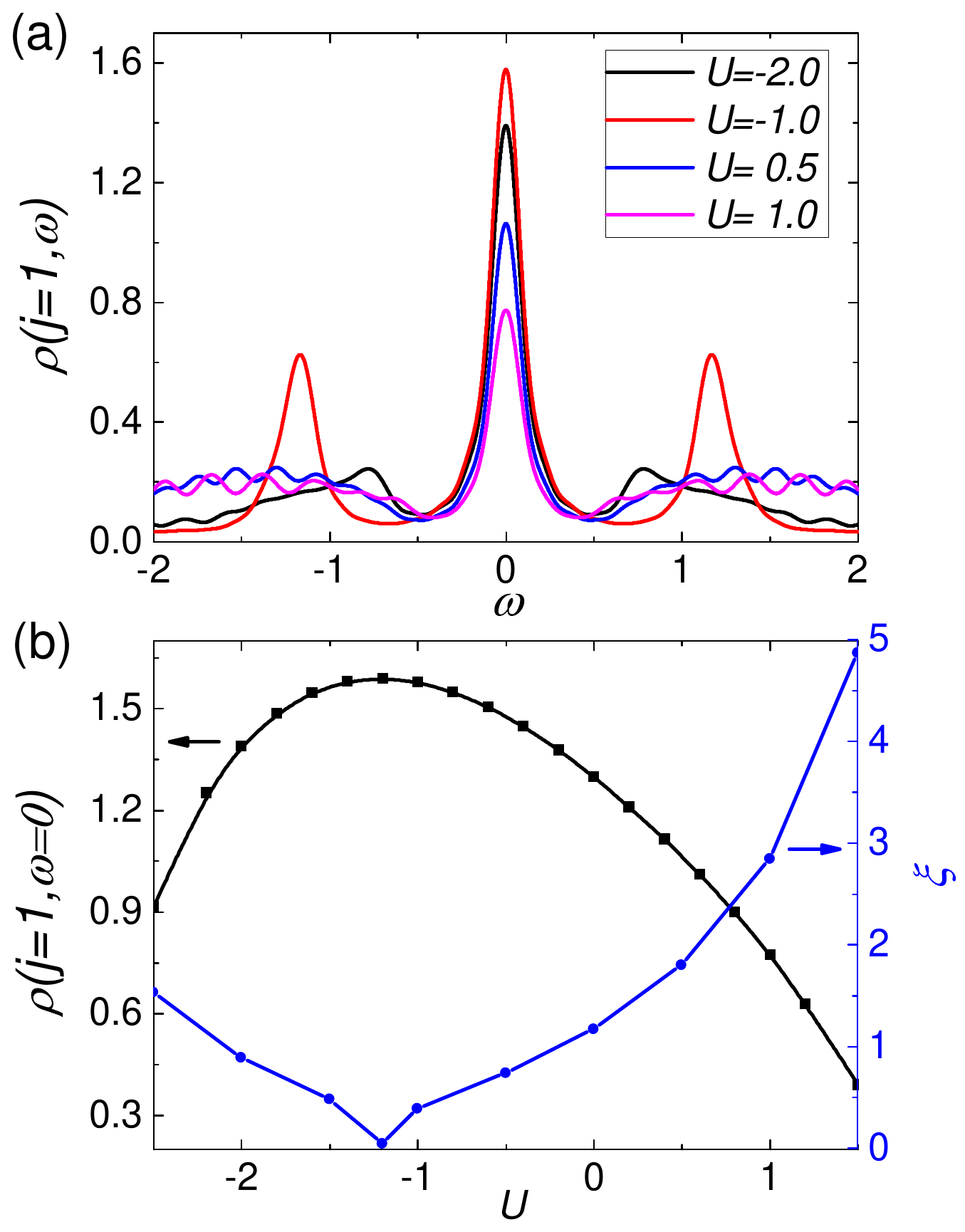}
\end{center}
 \caption{(Color online) (a) The LDOS as a function of energy at one end of the 1D interacting Kitaev chain for several different $U$'s.
          (b) The LDOS peak height at $\omega=0$ (black solid line) and the characteristic length $\xi$ (blue solid line) as functions of $U$. The values of $\xi$ are obtained by fitting an exponential decay function $e^{-(j-1)/\xi}$ to the data of $\rho(j,\omega=0)$
          The system parameters are $\Delta=0.4$ and $L=20$.
         }
 \label{Fig_LDOS}
\end{figure}

Figure~\ref{Fig_LDOS}(a) shows our numerical results of LDOS as a function of energy at one end of a finite chain, i.e. $\rho(1,\omega)$.  Prominent zero-bias LDOS peaks can be seen, which manifests the existence of MZM as a zero-energy boundary mode in the topological phase($-U_c<U<U_c$). By examining the peak height $\rho(1,\omega=0)$ as well as the decay of LDOS at zero energy $\rho(j,\omega=0)$, we find that the effect of the $U$ term on MZM is obviously not monotonic. For $\Delta=0.4$, the highest zero-energy LDOS peak together with shortest characteristic length occur at $U_p=-1.2$ as shown in Fig.~\ref{Fig_LDOS}(b). At $U=-1.2$, $\xi$ is vanishingly small indicating that MZM is exactly localized. We also perform the same calculations for the values of $\Delta$ varying from $0.1$ to $2.0$ and confirm that $U_p=-2(t-\Delta)$ is indeed the optimal value at which the MZM is maximally strengthened by interaction.
Contrary to previous results,~\cite{Ref_interaction_3} our results indicate that the decisive factor of $U$  is not its sign and the Majorana zero-energy peak is always weakened if $U$ is less or more than $U_p$, no matter whether the interaction is attractive or repulsive. Furthermore, the exactness of $U_p$ can be proven by the following discussion. Actually we find that $(\mu=0$, $U_p=-2(t-\Delta))$ corresponds to a frustration-free point at which the ground states can be given exactly as,~\cite{Ref_interaction_5}
\begin{equation} \label{gs}
    |\psi_0^\pm\rangle = \frac{1}{2^{L/2}}\prod_{j=1}^{L} (1\pm c_j^\dagger)|0\rangle.
\end{equation}
The exact form of the many-body MZM at $U_p$ can be written as sum of products of odd number of Majorana fermions $a_j$, $b_j$,
\begin{equation}
    \gamma = \sum_{j=1}^{L} a_j \prod_{k=1}^{j-1} (-ia_k b_k),
\end{equation}
where $a_j=c_j+c_j^\dagger$ and $b_j=-i(c_j-c_j^\dagger)$.  One can readily check that $\gamma$ is an exact zero mode satisfying $[H,\gamma]=0$ even for a finite chain. Substituting Eq.~(\ref{gs}) into Eq.~(\ref{green}) we find that $\rho(j,\omega)=(\delta_{j,1}+\delta_{j,L})\delta(\omega)$, which
rigorously prove the existence of exact zero modes with extremely localized contribution to the zero-energy LDOS at $U_p$.

\begin{figure}[ht]
\begin{center}
  \includegraphics[width=8.00cm, angle=0]{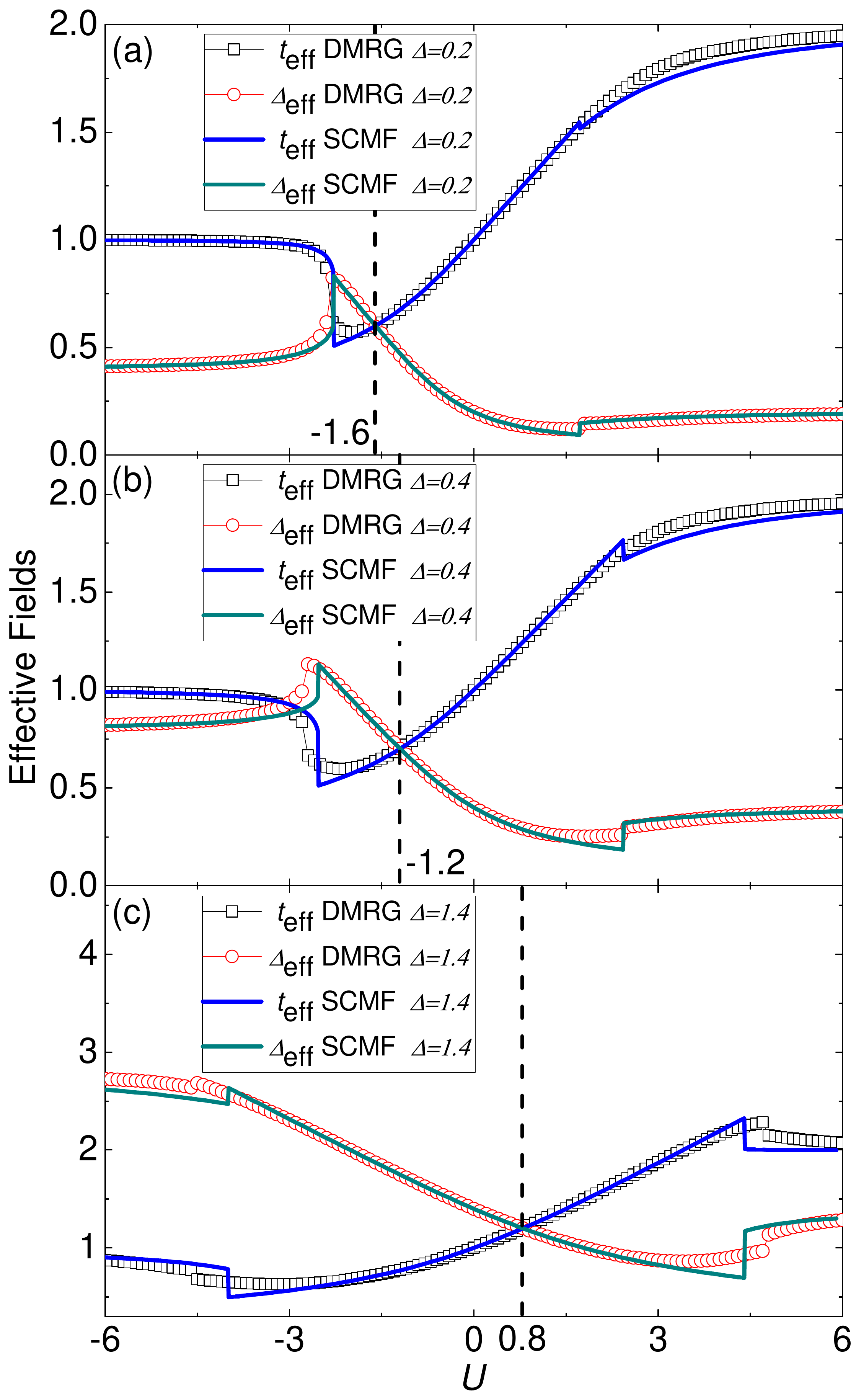}
\end{center}
 \caption{(Color online)
 The effective hopping integral $t_{\text{eff}}$ (blue line and black square)
 and the effective superconducting pairing potential $\Delta_{\text{eff}}$ (green line and red circle) as functions of $U$
 obtained by SCMF and DMRG for $\Delta=0.2$~(a), $0.4$~(b), $1.4$~(c). The positions of $U_p$ are emphasized using black dashed lines.
 $L=100$ for SCMF and $L=32$ for DMRG. }
 \label{Fig_Eff_Fields}
\end{figure}
To have a simple and intuitive understanding of the DMRG results, we next employ the mean field approximation to reduce the intractable interacting Kitaev chain into a more tractable effective noninteracting one. The SCMF Hamiltonian of Eq.~\eqref{Eq_Ham_00} is obtained after decoupling the quartic term in all three channels according to Wick's theorem,~\cite{Ref_U_Wick_decomposition}
\begin{align}\label{Eq_Ham_eff}
    H_{\text{MF}} =  & \sum_{j} \big[ - t_{\text{eff}}(j) c_{j}^{\dagger}c_{j+1}
                                       - \Delta_{\text{eff}}(j) ~ c_{j}^{\dagger}c_{j+1}^{\dagger} + h.c. \big] \nonumber \\
              & -\sum_{j}\mu_{\text{eff}}(j) (c_{j}^{\dagger}c_{j} - \frac{1}{2}),
\end{align}
where,
\begin{align}
        & t_{\text{eff}}(j)      =  t      + U  \langle c_{j+1}^{\dagger} c_{j} \rangle_{\text{MF}}, \label{teff}\\
        & \Delta_{\text{eff}}(j) =  \Delta + U  \langle c_{j}c_{j+1} \rangle_{\text{MF}}, \label{deff} \\
        & \mu_{\text{eff}}(j)    =  \mu - U  ( \langle n_{j-1}  \rangle_{\text{MF}} + \langle n_{j+1} \rangle_{\text{MF}} -1 ),
\end{align}
and $\left\langle\cdots\right\rangle_{\text{MF}}$ denotes the expectation value in the mean-field ground states.
Here the effective fields $t_{\text{eff}}(j)$, $\Delta_{\text{eff}}(j)$ and $\mu_{\text{eff}}(j)$ are calculated self-consistently.
As a comparison, the DMRG method is also employed to calculate the effective fields,
in other words, $\left\langle\cdots\right\rangle$ can also be implemented in terms of the DMRG ground states.

The site-independent effective fields obtained by SCMF and DMRG are showed in Fig.~\ref{Fig_Eff_Fields} at $\Delta=0.2$, $0.4$, $1.4$. In general, the variations of the effective fields as functions of $U$ from both methods are qualitatively compatible with each other. $t_{\text{eff}}$ and $\Delta_{\text{eff}}$ change continuously with $U$, except that there are two jumps at the phase boundaries $\pm U_c$, which signifies two phase transitions.
However, both the phase boundaries and the magnitudes of the discontinuities given by SCMF deviate from those given by DMRG. Furthermore we also perform DMRG calculations at various chain length $L$, and find obvious finite-size effect on the phase boundaries. On the other hand, for $U$ in the topological region especially at the vicinity of $U_p$, the SCMF results agree quantitatively well with the DMRG results even for finite chain length as shown in Fig.~\ref{Fig_Eff_Fields}. This finding can be understood as follows: (i) We find an intersection point denoted by $U_i$ where $t_\text{eff}$ and $\Delta_\text{eff}$ intersect with each other. Since $t_\text{eff}=\Delta_\text{eff}$ and $\mu_\text{eff}=0$ (not shown in the figure) at $U_i$, we obtain
\begin{equation} \label{expectation}
    \langle c_{j+1}^\dagger c_j\rangle_\text{MF}=-\langle c_j c_{j+1}\rangle_\text{MF}=1/4,
\end{equation}
based on the explicit ground state~\cite{Greiter} of the SCMF Hamiltonian (\ref{Eq_Ham_eff}) at these special parameters. Substituting Eq.~(\ref{expectation}) into Eqs.~(\ref{teff}), (\ref{deff}) and then solving $t_\text{eff}=\Delta_\text{eff}$, we have $U_i=-2(t-\Delta)=U_p$. (ii) One can further see that the ground state of the interacting Kitaev chain at $U_p$ as given in Eq.~(\ref{gs}) is actually the same as that of the noninteracting Kitaev chain at the special parameters $t_\text{eff}=\Delta_\text{eff}$ and $\mu_\text{eff}=0$.~\cite{Greiter} And thus the effective fields calculated by SCMF match accurately with those by DMRG  at $U_p$ even for finite chain size, as shown in Fig.~\ref{Fig_Eff_Fields}

When $U$ changes away from $U_p$ in the topological region, the difference between $t_{\text{eff}}$ and $\Delta_{\text{eff}}$ enlarges while $\mu_\text{eff}$ is always zero.  Such behavior of the three effective fields $t_\text{eff}$, $\Delta_\text{eff}$ and $\mu_\text{eff}$ as functions of $U$, especially $t_{\text{eff}}=\Delta_{\text{eff}}$ and $\mu_{\text{eff}}=0$ at $U=U_p$, explains the DMRG results of energy spectrum, entanglement entropy and LDOS: (i) From the quasiparticle excitation spectrum of the SCMF Hamiltonian $\epsilon(k)=\sqrt{(2t_{\text{eff}} \cos k+\mu_{\text{eff}})^2 + (2\Delta_{\text{eff}} \sin k)^2}$, the excitation gap has a local maximum at $U=U_p$ as seen in Fig.~\ref{Fig_En}. (ii) The variation of $\xi$ with $U$ as shown in Fig.~\ref{Fig_LDOS} can also be qualitatively described by the equation $\xi=\ln^{-1}[{(t_\text{eff}+\Delta_\text{eff})}/{|t_\text{eff}-\Delta_\text{eff}}|]$ which is valid for noninteracting
Kitaev chain at $\mu_{\text{eff}}=0$. (iii) The lift of the ground state degeneracy is proportional to $e^{-L/\xi}$. $t_{\text{eff}}=\Delta_{\text{eff}}$ at $U_p$ leads to vanishing $\xi$ and exact two-fold degeneracy even for a finite-size chain, which explains the minimum of entropy at $U_p$ as shown in Fig.~\ref{Fig_En} and the highest LDOS peak in Fig.~\ref{Fig_LDOS}.

In summary, we have investigated the interacting Kitaev chain at half filling. Three self-dual points of the Hamiltonian are given, two of which correspond to the phase boundaries. The third one, $U_p$, is located in the topological region, whose significance is revealed by calculating the low-energy excitation spectrum, the entanglement entropy and the LDOS etc using the DMRG method as well as analytic derivation. The effect of interaction on Majorana zero modes can be described by the mean field approximation qualitatively. Furthermore we find that the effective fields calculated by the SCMF method agree quantitatively with those by the DMRG method at the vicinity of $U_p$.

This work was supported by the National Natural Science Foundation of China under Grant No 11274379, and the Research Funds of Renmin University of China under Grant No 14XNLQ07.

\end{document}